\documentclass[twocolumn]{aastex631}

\graphicspath{{./}{figures/}}

\usepackage{color}
\usepackage{amsmath}
\usepackage{xcolor}
\usepackage{mathrsfs}
\usepackage{amssymb}
\usepackage{natbib}
\usepackage{gensymb}
\usepackage{xspace}
\usepackage{subfigure}
\usepackage{placeins}
\usepackage{graphicx}
\usepackage{booktabs}
\usepackage{float}

\newcommand{\Rj}{\ensuremath{R_{\rm{Jup}}}\xspace}
\newcommand{\Mj}{\ensuremath{M_{\rm{Jup}}}\xspace}

\newcommand{\Lsun}{L_\odot}
\newcommand{\Teff}{\ensuremath{T_\mathrm{eff}}\xspace}
\newcommand{\logg}{\ensuremath{\log{g}}\xspace}
\newcommand{\lbollsun}{\ensuremath{\log(L_\mathrm{bol}/\Lsun)}}
\newcommand{\lbollsunba}{\ensuremath{\log(L_\mathrm{bol, 1}/\Lsun)}}
\newcommand{\lbollsunbb}{\ensuremath{\log(L_\mathrm{bol, 2}/\Lsun)}}
\newcommand{\lbol}{\ensuremath{L_\mathrm{bol}}}

\newcommand{\logkzz}{\log\ensuremath{K_\mathrm{zz}}\xspace}

\newcommand{\logkzzbau}{\log(\ensuremath{K_\mathrm{zz,1} / \mathrm{cm^2 s^{-1}}})\xspace}
\newcommand{\logkzzbbu}{\log(\ensuremath{K_\mathrm{zz,2} / \mathrm{cm^2 s^{-1}}})\xspace}

\newcommand{\teffba}{\ensuremath{T_\mathrm{eff,1}}\xspace}
\newcommand{\teffbb}{\ensuremath{T_\mathrm{eff,2}}\xspace}
\newcommand{\loggba}{\ensuremath{\log{g}_{,1}}\xspace}
\newcommand{\loggbb}{\ensuremath{\log{g}_{,2}}\xspace}

\newcommand{\caltech}{Department of Astronomy, California Institute of Technology, Pasadena, CA 91125, USA}
\newcommand{\gps}{Division of Geological \& Planetary Sciences, California Institute of Technology, Pasadena, CA 91125, USA}
\newcommand{\ucsc}{Department of Astronomy \& Astrophysics, University of California, Santa Cruz, CA95064, USA}

\newcommand{\jpl}{Jet Propulsion Laboratory, California Institute of Technology, 4800 Oak Grove Dr.,Pasadena, CA 91109, USA}
\newcommand{\ucsd}{Department of Astronomy \& Astrophysics,  University of California, San Diego, La Jolla, CA 92093, USA}

\newcommand{\carnegiew}{Earth and Planets Laboratory, Carnegie Institution for Science, Washington, DC, 20015}

\newcommand{\stsci}{Space Telescope Science Institute, 3700 San Martin Drive, Baltimore, MD 21218, USA}

\newcommand{\jhu}{Department of Physics and Astronomy, Johns Hopkins University, Baltimore, MD, USA}

\newcommand{\curL}{\mathcal{L}}

\begin{document}

\title{Atmospheric abundances and bulk properties of the binary brown dwarf Gliese 229 Bab from JWST/MIRI spectroscopy}

\correspondingauthor{J. Xuan}
\email{wxuan@caltech.edu}

\author[0000-0002-6618-1137]{Jerry W. Xuan}
\affiliation{\caltech}

\author[0000-0002-3191-8151]{Marshall D. Perrin}
\affiliation{\stsci}

\author{Dimitri Mawet}
\affiliation{\caltech}
\affiliation{\jpl}

\author[0000-0002-5375-4725]{Heather A. Knutson}
\affiliation{\gps}

\author[0000-0003-1622-1302]{Sagnick Mukherjee}
\affiliation{\ucsc}
\affiliation{\jhu}

\author[0000-0003-0097-4414]{Yapeng Zhang}
\affiliation{\caltech}

\author[0000-0002-9803-8255]{Kielan K. W. Hoch}
\affiliation{\stsci}

\author[0000-0003-0774-6502]{Jason J. Wang}
\affiliation{Center for Interdisciplinary Exploration and Research in Astrophysics (CIERA) and Department of Physics and Astronomy, Northwestern University, Evanston, IL 60208, USA}

\author[0000-0001-9164-7966]{Julie Inglis}
\affiliation{\gps}

\author[0000-0003-0354-0187]{Nicole L. Wallack}
\affiliation{\carnegiew}

\author[0000-0003-2233-4821]{Jean-Baptiste Ruffio}
\affiliation{\ucsd}

\begin{abstract}

We present JWST/MIRI low-resolution spectroscopy ($4.75-14~\mu$m) of the first known substellar companion, Gliese 229 Bab, which was recently resolved into a tight binary brown dwarf. Previous atmospheric retrieval studies modeling Gliese 229 B as a single brown dwarf have reported anomalously high carbon-to-oxygen ratios (C/O) of $\approx1.1$ using $1-5~\mu$m ground-based spectra. Here, we fit the MIRI spectrum of Gliese 229 Bab with a two-component binary model using the Sonora Elf Owl grid and additionally account for the observed $K$ band flux ratio of the binary brown dwarf. Assuming the two brown dwarfs share the same abundances, we obtain $\rm C/O=0.65\pm0.05$ and $\rm [M/H]=0.00^{+0.04}_{-0.03}$ as their abundances ($2\sigma$ statistical errors), which are fully consistent with the host star abundances. We also recover the same abundances if we fit the MIRI spectrum with a single brown dwarf model, indicating that binarity does not strongly affect inferred abundances from mid-infrared data when the $\Teff$ are similar between components of the binary. We measure $\Teff=900^{+78}_{-29}~$K and $\Teff=775^{+20}_{-33}~$K for the two brown dwarfs. We find that the vertical diffusion coefficients of $\logkzz \approx4.0$ are identical between the two brown dwarfs and in line with $\logkzz$ values inferred for isolated brown dwarfs with similar $\Teff$. Our results demonstrate the power of mid-infrared spectroscopy in providing robust atmospheric abundance measurements for brown dwarf companions and by extension, giant planets. 

\end{abstract}

\section{Introduction} \label{sec:intro}
Brown dwarf companions ($m\approx13-72~\Mj$) to stars provide excellent tests of substellar atmospheric and evolutionary models, and serve as a key bridge to understanding the population of directly imaged giant exoplanets. From the powerful combination of \textit{Hipparcos}-\textit{Gaia} astrometry \citep{brandt_hipparcos-gaia_2018, kervella_stellar_2019, xuan_Evidence_2020} and long-term radial velocity and imaging observations, a growing subset of brown dwarf companions now have their dynamical masses measured \citep[e.g.][]{brandt_229B_2019, brandt_precise_2019, Franson2022_HD984, Franson2023_HIP, Franson2023_PZTel, Rickman2024}. The dynamical masses can be compared with bolometric luminosity and stellar age measurements to benchmark substellar evolutionary models \citep[e.g.][]{Zapatero2004, Stassun2006, Dupuy2014, Dupuy2017}. Such studies have revealed a number of discrepancies between models and observations, including a sample of older brown dwarf companions that appear under-luminous given their measured masses (typically $>2$ Gyr; \citealt{Cheetham2018, brandt_229B_2019, Bowler2021_HD47127}), and a number of young companions that are over-luminous for their masses \citep[e.g.][]{Dupuy2009, Dupuy2014, Brandt2021}. It is crucial to investigate the reason behind these discrepancies, as the same models are used to estimate model-dependent masses for the vast majority of imaged planets and brown dwarf companions that lack direct mass measurements, for instance due to prohibitively long orbital periods.

Dynamical masses also provide crucial information to aid atmospheric studies, which are often under-constrained and display complex degeneracies between surface gravity, clouds, and metallicity \citep[e.g.][]{ZJZhang2021, Xuan2022, Landman2023, Balmer2024_HD136164}. The introduction of dynamical mass priors in atmospheric studies can help to exclude nonphysical parts of the parameter space, and have become widely adopted whenever available \citep[e.g.][]{Zhang2023, Nasedkin2024, Hsu2024}. Brown dwarf companions that have both dynamical masses and precise spectroscopic observations sensitive to multiple molecular species provide the most stringent tests of substellar models. For example, numerous studies using high-resolution ground-based spectroscopy have shown that companions with $m\gtrsim40~\Mj$ are chemically homogeneous with their host stars, as expected for objects that formed via direct gravitational collapse from the same disk or cloud \citep[e.g.][]{Xuan2022, wang_Retrieving_2022, Hsu2024, Costes2024, Xuan2024}. Recently, \citet{Xuan2024b} also showed that this trend extends to the $\sim10-30~\Mj$ regime, implying that gravitational instabilities can form systems with a broad range of masses and mass ratios (see also \citealt{Hoch2023}).  

Gliese 229 B is one of the closest brown dwarf systems (5.761 parsec; \citealt{Gaia2022_DR3}) \citep{Nakajima1995,Oppenheimer1995}, orbiting the M1V star Gliese 229A on a highly eccentric orbit ($e\approx0.8$) with semi-major axis of $33$ AU. \citet{Brandt2021} measured a precise dynamical mass of $71.4\pm0.6~\Mj$ for Gliese 229 B, which is unusually high given the low luminosity of the source. While Gliese 229 B has been studied as a benchmark single brown dwarf for nearly three decades, \citet{Xuan2024gliese} recently used VLTI/GRAVITY and VLT/CRIRES+ to show that Gliese 229 B is in fact comprised of two brown dwarfs, Ba and Bb, orbiting each other on a tight orbit with $a\approx0.042$ AU (or orbital period of 12.1 days).\footnote{See also the complementary evidence for binarity presented in \citet{Whitebook2024}.} The properties of Gliese 229 BaBb resolve previous tensions between mass and luminosity and recast the system as a benchmark brown dwarf binary in a triple system. 

There have been numerous atmospheric studies on Gliese 229 B over the years using optical and near-infrared spectrum from $0.6-2.5~\mu$m, and in some cases, $0.6-5~\mu$m. The early studies mainly matched the spectrum with model grids or performed by-eye identification of absorption features \citep[e.g.][]{marley1996, Geballe1996, Noll1997,Oppenheimer1998,Saumon2000,Leggett2002}. To summarize, these studies constrain the $\Teff$ between about 850 to 1100 K, find evidence of disequilibrium chemistry from the over-abundance of CO, but disagree strongly on the surface gravity, which ranges from $\logg=3.5$ to $\logg=5.3$. Two of these studies attempted to measure the metallicity of Gliese 229 B, and find sub-solar values between $-0.5$ to $-0.3$ dex \citep{Saumon2000, Leggett2002}, but noted that the metallicity is degenerate with the poorly constrained surface gravity. 

There have also been two atmospheric retrievals published using the same near-infrared spectra of Gliese 229 B \citep{calamari_Atmospheric_2022, howe_GJ_2022}. Both of these retrieval studies find a solar metallicity for Gliese 229 B, unlike the earlier studies. However, both studies also found an elevated $\rm C/O \approx1.1$, which is unexpected given the nearly solar $\rm C/O=0.68\pm0.12$ of the primary star \citep{Nakajima2015}. C/O values exceeding 1 have been reported for a number of T dwarfs using low-resolution, near-infrared spectroscopy \citep[e.g.][]{Zalesky2019, Zalesky2022, Gaarn2023}, and is usually thought to be an observational bias caused by either data systematics or model inaccuracies (or both), though the exact reason is unclear \citep{Gaarn2023, Calamari2024}. While there could be some T dwarfs with actual $\rm=C/O > 1$, such objects should be quite rare given the paucity of FGK stars with $\rm=C/O > 1$ in the solar neighborhood \citep[e.g.][]{brewer_SPECTRAL_2016}. Finally, the discovery that Gliese 229 B is a binary brown dwarf \citep{Xuan2024gliese} naturally raises the question of what effect, if any, its binarity might have on atmospheric retrieval studies.

Although most previous atmospheric studies of Gliese 229 B treated this companion as a single object, \citet{Howe2023} tried fitting a two-component grid model to its $1-5\mu$m spectra. However, due to limitations of the models and the limited quality and wavelength coverage of the data, their binary fits produced significant residuals and were not statistically preferred over the single-brown dwarf retrievals in \citet{howe_GJ_2022}. The properties inferred for the brown dwarf binary from \citet{Howe2023} are also inconsistent with the mass ratio and flux ratio measured by \citet{Xuan2024gliese}.

In this paper, we present JWST Mid Infrared Instrument (MIRI) low-resolution spectroscopy of Gliese 229 B from $4.75-14$~$\mu$m. Informed by the results from \citet{Xuan2024gliese}, we model the spectrum using a binary brown dwarf model and estimate the effective temperature, surface gravity, and vertical mixing rate for each component. In addition, we provide updated measurements of the binary's atmospheric metallicity and C/O ratio.

\begin{deluxetable}{ll}[t!]
\tablecaption{Fitted Parameters and Priors}\label{tab:param_prior}
\tabletypesize{\small}
\tablehead{Parameter & Prior}
\startdata
Physical parameters \\
\hline
Total Mass ($\Mj$) & $\mathcal{N}(71.3, 0.5)$   \\
$q$ & $\mathcal{N}(0.91, 0.05)$   \\
$\teffba$ ($K$) & $\mathcal{U}(575, 1200)$  \\
$\teffbb$ ($K$) & $\mathcal{U}(575, \teffba)$  \\
$\loggba$ & $\mathcal{U}(4.2, 5.5)$  \\
$\loggbb$ & $\mathcal{U}(\loggba, 5.5)$  \\
C/O & $\mathcal{U}(0.23, 1.15)$  \\
$\rm{[M/H]}$ & $\mathcal{U}(-1.0, 1.0)$ \\
$\logkzzbau$ & $\mathcal{U}(2.0, 9.0)$ \\
$\logkzzbbu$ & $\mathcal{U}(2.0, 9.0)$ \\
\hline
Nuisance parameters \\
\hline
$10^b$ & $\mathcal{U}( \rm{0.01\times min(\epsilon_i^2), 100\times max(\epsilon_i^2)} )$ \\
$w_0$ ($\mu$m) & $\mathcal{U}(-0.15, 0.15)$ \\
$w_1$ & $\mathcal{U}(0.96, 1.04)$ \\
$w_2$ ($\mu$m$^{-1}$) & $\mathcal{U}(-0.003, 0.003)$ \\
$r0$ & $\mathcal{U}(-150, 0)$  \\
$r$ ($\mu$m$^{-1}$) & $\mathcal{U}(10, 35)$  \\
\enddata
\tablecomments{We use $_1$ to denote Gliese 229 Ba, the primary brown dwarf, and $_2$ to denote Gliese 229 Bb, the secondary brown dwarf. In the single brown dwarf model, the priors adopted are the same as those for the $_1$ component. $\mathcal{U}$ stands for a uniform distribution, with two numbers representing the lower and upper boundaries. $\mathcal{N}$ stands for a Gaussian distribution, with numbers representing the mean and standard deviation. For the $10^{b}$ error inflation factor, $\epsilon_i$ refers to the data uncertainties (see Eq.~\ref{eq:einflate}).}
\end{deluxetable}

\section{Observations and Data Reduction} \label{sec:obs}
We observed Gliese 229 B using JWST's Mid Infrared Instrument (MIRI; \citealt{Rieke2015}) on UT 2023 December 13 (GO3762, PI: Xuan). The observations were carried out with the low-resolution
spectrometer (LRS; \citealt{Kendrew_2015}) in fixed slit mode ($0\farcs51\times4\farcs7$). We obtained spectroscopy from $5–14~\mu$m with an average resolving power of $\sim100$. 

At the start of the observation, a target acquisition exposure was taken on the star Gliese 229 to measure and correct for observatory pointing uncertainties, and an offset move was performed to position the companion Gliese 229 B within the LRS slit. A target acquisition confirmation exposure was taken (F560W filter, 16.6 seconds) to allow precise confirmation measurement of the achieved position of Gliese 229 B within the slit. Finally, two spectral exposures were taken with the LRS prism. We used the typical two-point ``along slit nod'' dither \citep{Gordon2015}, and observed for a total exposure time of 571.66 s in FASTR1 readout mode using 25 groups per integration. The MIRI data used in this paper can be found in MAST: \dataset[10.17909/950e-3e83]{http://dx.doi.org/10.17909/950e-3e83}.

\subsection{Data Reduction and Forward Modeling Host Star Contamination}\label{sec:data_red}

We reduced the data using the JWST pipeline (version $1.14.0$ and CRDS context $\rm{jwst1255.pmap}$), plus additional custom steps to clean bad pixel outliers and to subtract the modest amount of contamination from host starlight within the slit. 

Speckle contamination arises from the host star point spread function (PSF) diffraction pattern. Light from the star's PSF wings that enters the LRS slit accounts for approximately $10\%$ of the flux at the companion location. To remove stellar contamination, we forward model the off-axis host star PSF with WebbPSF \citep{Perrin2012SPIE.8442E..3DP,Perrin2014SPIE.9143E..3XP}. The details of forward modeling and subtraction procedure are provided in Appendix~\ref{app:data_reduction}.

After subtracting the stellar PSF, the two science observation nods are subtracted from one another to remove the observatory thermal background. We extracted independent sets of spectra from the two spectral traces. A key performance metric for the forward modeling process is the consistency of the spectra measured for each of the two nods. Without subtracting the host star PSF, the spectra from the two nods differ by $\approx 10\%$; after subtraction they agree to $\approx 2-3\%$.

\subsection{Extracted Spectrum}\label{sec:spec_extract}
\begin{figure*}[t!]
    \centering
\includegraphics[width=0.95\linewidth]{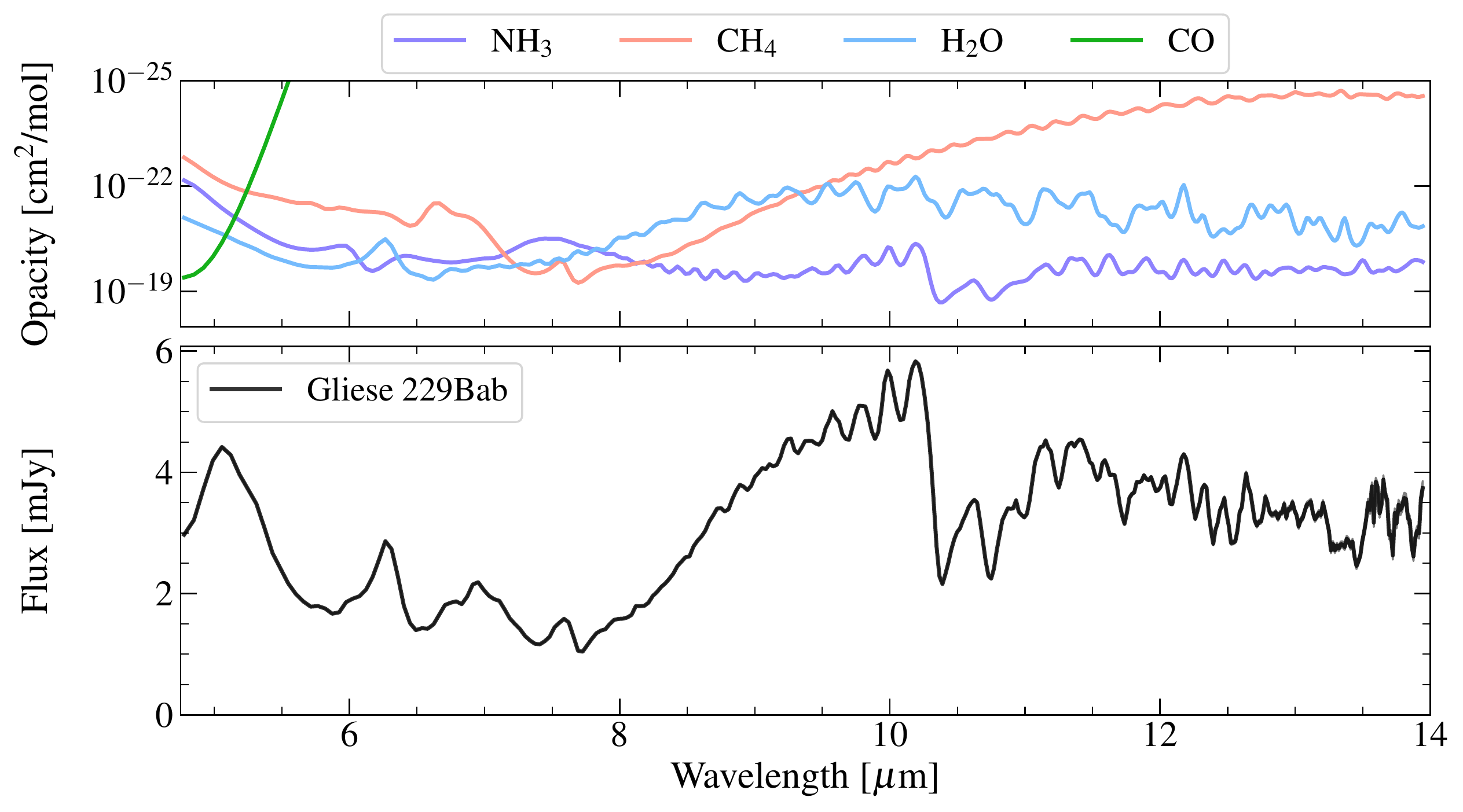}
    \caption{Top panel: Opacities of the major molecular absorbers in the mid-infrared spectrum of Gliese 229 Bab. The units are $\rm{cm}^2$ per molecule. Note that the y-axis is inverted. Bottom panel: MIRI LRS spectrum of Gliese 229 Bab in black. The error bars are shown as shaded regions around the spectrum, but are not visible by eye except for the reddest wavelengths.}
    \label{fig:data_opas}
\end{figure*}

After obtaining the nod-subtracted 2D spectrum, we proceed to extract the 1D spectrum using the JWST Level 3 pipeline. Specifically, we perform a box extraction centered on the spectral traces using the \texttt{extract\_1d} function in the pipeline, with a box width of 8 pixels. The width is chosen to fully enclose the core of the spectral trace, while minimizing the included background area. Aperture correction is taken into account based on our box width, and the pipeline combines the two nods into a final spectrum. The extracted spectrum of Gliese 229 Bab is shown in Fig.~\ref{fig:data_opas}, along with opacities of the major relevant molecular sources. 

The nominal wavelength range for MIRI LRS is $5-14~\mu$m. Our spectrum for Gliese 229 Bab has an average S/N of $\approx400$ per point compared to the background noise, although systematics from subtracting the stellar speckles limit our effective S/N to $\approx30-50$ given the $2-3\%$ difference between nods. While the $10-14~\mu$m region suffers more strongly from an inaccurate wavelength solution, this can be corrected with a second order polynomial (see \S~\ref{sec:reslambda} and Fig.~\ref{fig:wcorr}). Furthermore, after carrying out initial fits to the $5-14~\mu$m data, we find that our best-fit models can be extended blueward to $4.75~\mu$m and still fit the data well. Extending to $4.75~\mu$m allows us to capture the end of the mid-infrared CO absorption band. Therefore, we adopt a wavelength range of $4.75-14~\mu$m for the MIRI spectrum presented in this paper. 

\section{Spectral analysis} \label{sec:spec_analysis}

\subsection{Resolving power and wavelength correction}\label{sec:reslambda}
First, we describe our treatment of two important aspects of the data, the variable resolving power, and the imperfect wavelength solution available at the time of analysis. We account for both these effects using additional nuisance parameters.

The resolving power ($R = \lambda/\Delta\lambda$) of MIRI LRS increases with wavelength in a nearly linear manner. In this work, we account for this by fitting a linear relation between $R$ and the data wavelengths, $\lambda$

\begin{equation}
    R_\lambda = r_0 + r\lambda
\end{equation}

where $r$ and $r_0$ are free parameters. In each iteration of the fit, the model spectrum is convolved with a variable Gaussian kernel whose standard deviation $\sigma$ is given by 

\begin{equation}
    \sigma = \frac{\lambda / R_\lambda} {2\sqrt{2\log{2}}}
\end{equation}

Second, our initial fits show inaccuracies in the wavelength solution, especially at longer wavelengths. We include a quadratic wavelength correction to improve the wavelength solution. The corrected data wavelengths ($\lambda^\prime$) are given by

\begin{equation}
    \lambda^\prime = w_0 + w_1\lambda + w_2\lambda^2
\end{equation}

where $\lambda$ is the wavelength from the pipeline, and $w_0$, $w_1$, and $w_2$ are coefficients of the polynomial that we fit for. In practice, we fit the parameters for $R_\lambda$ and $\lambda^\prime$ once using a single brown dwarf model and subsequently adopt the best-fit values for subsequent fits.

\subsection{Model fitting with Elf Owl}
We fit the MIRI LRS data of Gliese 229 B using the Sonora Elf Owl models \citep{Mukherjee2024}. The Elf Owl grid includes five parameters: the effective temperature ($\Teff$), surface gravity ($\logg$), vertical diffusion coefficient ($\logkzz$), carbon-to-oxygen ratio (C/O), and bulk metallicity ($\rm{[M/H]}$). While the models are cloudless, we do not expect cloud opacity to affect the near or mid infrared spectra of late T dwarfs \citep[e.g.][]{Line2017, Suarez2022_L2silicate, Mukherjee2024}. We note that alternative cloudless models for brown dwarfs exist, and some recent ones include ATMO 2020, Sonora Bobcat, and Sonora Cholla \citep{Phillips2020, marley_Sonora_2021, karalidi_Sonora_2021}. However, ATMO 2020 assumes solar abundances, whereas Bobcat varies metallicity and C/O but does not account for disequilibrium chemistry. In addition, Cholla varies the $\logkzz$ but fixes the abundances to solar. While we only use Elf Owl in this paper, it would be informative to compare how well different models fit the MIRI data of Gliese 229 B in future studies.

Using JWST/NIRSpec spectroscopy, \citet{Beiler2024b} showed that substellar models including Sonora Elf Owl under-estimate the CO$_2$ abundance in brown dwarf atmospheres due to inaccurate treatments of disequilbrium chemistry. On the other hand, the non-detection of PH$_3$ suggests an over-prediction of the PH$_3$ abundance. Our MIRI LRS data are not sensitive to the strongest features from CO$_2$ and PH$_3$ at approximately $4.2-4.4~\mu$m.

As Gliese 229 B has been resolved into a binary brown dwarf \citep{Xuan2024gliese}, our fiducial model is a two-component binary model that allows different $\Teff$, $\logg$, and $\logkzz$ values for each brown dwarf. Given the limiting resolution of our MIRI LRS data, we assume the C/O and $\rm{[M/H]}$ to be identical between the two brown dwarfs in our fiducial model. However, we also fit models with two distinct sets of C/O and $\rm{[M/H]}$ for each brown dwarf to test this assumption. We parameterize the component masses using the mass ratio ($q$) and total mass. The component masses along with the individual $\logg$ are used to compute the radii, which scales the model flux along with the distance. The distance to the source is fixed at 5.761 parsec based on the Gaia DR3 parallax measurement \citep{Gaia2022_DR3}.

For the total mass and mass ratio, we adopt Gaussian priors from \citet{Xuan2024gliese} (see Table~\ref{tab:param_prior}). From physical considerations of energy conservation and degeneracy pressure, we always expect the more massive brown dwarf to be hotter and have a higher surface gravity than the less massive component. Thus, we put priors on the secondary brown dwarf such that its \Teff and \logg are always lower than that of the primary component.

In addition to nuisance parameters for the resolving power and wavelength correction (\S~\ref{sec:reslambda}), we fit an error inflation term $b$ following \citet{line_Uniform_2015}, which scales the pipeline errors ($\epsilon$) such that 

\begin{equation}\label{eq:einflate}
    \epsilon^\prime=\sqrt{\epsilon + 10^{b}}
\end{equation}
where $\epsilon^\prime$ are the adopted errors for the log likelihood calculation. Fitting the error inflation term allows us to account for systematic uncertainties from the speckle subtraction (\S~\ref{sec:spec_extract}) and model uncertainties. In total, we have 10 physical parameters and 6 nuisance parameters for the binary model, which are summarized in Table~\ref{tab:param_prior}. In addition to the binary model, we also consider a single brown dwarf model to assess whether ignoring binarity changes the atmospheric composition measurements from the MIRI data.

\subsection{$K$ band flux ratio from GRAVITY}
To further constrain the atmospheric models, we fit the $K$ band ($2.025-2.15$ $\mu$m) flux ratio ($K_{\rm Bb} / K_{\rm Ba}$) of $0.50\pm0.03$ between Gliese 229 Ba and Bb measured by GRAVITY \citep{Xuan2024gliese}. To do so, we integrate the model spectra for each brown dwarf from $2.025-2.15$ $\mu$m and compute the model flux ratio between Bb and Ba. In summary, the joint likelihood is

\begin{equation}
    ln\curL = -\frac{1}{2}(\chi^2_{MIRI}+ \chi^2_{K_{\rm Bb} / K_{\rm Ba}})
\end{equation}

\begin{figure*}[t!]
    \centering
    \includegraphics[width=0.95\linewidth]{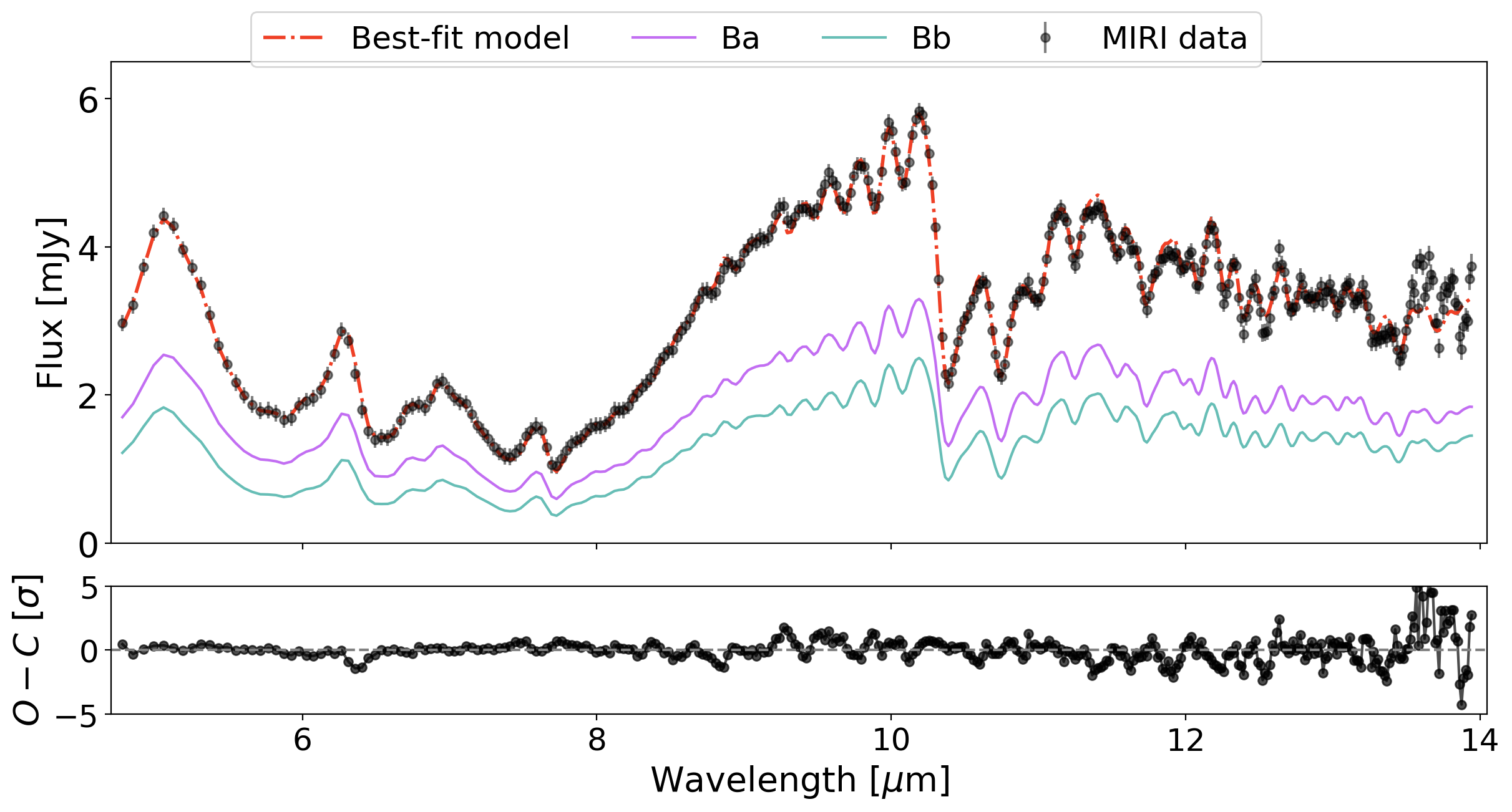}
    \caption{Top panel: MIRI LRS spectrum of Gliese 229 Bab in black circles. The error bars shown have been inflated by the best-fit error inflation factor. The red dashdot curve shows the best-fit model spectrum, which is the addition of the individual component spectra, Ba (purple), and Bb (teal). Bottom panel: residuals of the fit (data-model) plotted in units of $\sigma$ differences. Overall, the data are well-fit by the model ($<2\sigma$ residuals), except for the wavelength region past $12\mu$m where the pipeline calibration is still work in progress.}
    \label{fig:model_data}
\end{figure*}

\subsection{Nested sampling}
To sample the posteriors, we use the nested sampling package \texttt{dynesty} \citep{speagle_DYNESTY_2020}. We adopt 1000 live points and stop sampling when the estimated contribution of the remaining prior volume to the total evidence is $<1\%$. 

\section{Results} \label{sec:results}
Here, we discuss the key results of our spectral fits. We focus on the results of the fiducial model, i.e. the binary brown dwarf model that assumes the same abundances for Gliese 229 Ba and Bb. The best-fit fiducial model is shown in Fig.~\ref{fig:model_data}, and the retrieved parameters are listed in Table~\ref{table:results}. 

The exquisite spectro-photometric precision of MIRI spectrum produces very small uncertainties for most parameters, even after including the error inflation term in our fits. In all cases, the $68\%$ confidence intervals of the posterior estimates are smaller than the grid spacing of the model parameters. To be conservative, we report the $95\%$ confidence intervals for all parameters in this work. Grid model fits yielding small uncertainties have been observed in previous work \citep[e.g.][]{ZJZhang2021, Petrus2024}, and half the grid spacing could be a more conservative estimate of the model fitting uncertainties. In the relevant parameter space of Elf Owl we are using, half the grid spacing would correspond to uncertainties of 25 K for $\Teff$, $0.125$ dex for $\logg$, 0.11 for C/O, 0.25 dex for $\rm{[M/H]}$, and 1 dex for $\logkzz$.

In reality, other sources of uncertainties could also be important, such as those arising from model or data systematics. From inter-comparing substellar model grids, \citet{Lueber2023} showed that different models disagree strongly on the inferred surface gravity for the same brown dwarf, while \citet{Sanghi2023} highlight issues with the BT-Settl and ATMO 2020 atmospheric models \citep{Allard2012, Phillips2020} in yielding accurate $\Teff$ at the M/L transition and for L dwarfs. For Gliese 229 Bab, our precise knowledge of their dynamical masses significantly benefits the spectral fits in this regard. For high-contrast companions, other sources of error include correlated noise from the wavelength-dependence of the stellar PSF \citep[e.g.][]{Greco2016, Xuan2022, Nasedkin2023}.

\subsection{Resolving power and wavelength correction}
We include the parameters for resolving power and wavelength correction (\S~\ref{sec:reslambda}) in an initial single brown dwarf fit. The parameters are well-constrained, and we subsequently adopt the best-fit values for these parameters in our final binary fit. We tested that freely fitting for these nuisance parameters in the binary brown dwarf model does not affect the results of the binary model. 

Our retrieved resolving power as a function of wavelength is shown in Fig.~\ref{fig:rdraws}, and quantitatively agrees with an $R$ that is set by the size of the diffraction-limited PSF \citep{Beiler2023}. Accounting for the quadratic wavelength correction visually improves the fit for several absorption features (see Fig.~\ref{fig:wcorr}) and is statistically favored at the $13\sigma$ level. This suggests that there is room for additional improvements to the default MIRI LRS wavelength solution from the JWST pipeline.

\subsection{Bulk properties}
We find $\Teff=900^{+78}_{-29}~$K, $\logg=5.15^{+0.15}_{-0.04}$, $R=0.81^{+0.05}_{-0.12}~\Rj$ for the primary brown dwarf, Ba, and $\Teff=775^{+20}_{-33}~$K, $\logg=5.07^{+0.04}_{-0.11}$, $R=0.85^{+0.12}_{-0.05}~\Rj$ for the secondary brown dwarf, Bb ($2\sigma$ uncertainties, see Fig.~\ref{fig:corner_bulk}). We note that these posteriors show long tails which result from covariances between the parameters. The radius and \Teff of Ba and Bb are correlated; for example, a larger radius of Ba correlates with a smaller radius for Bb to match the total flux. This correlation between radius also explains the correlation between $\logg$ of each brown dwarfs. Overall, our measured values for \Teff, \logg, and radius are consistent at the $1\sigma$ level with values from \citet{Xuan2024gliese}, who derived bulk properties using ATMO 2020 evolutionary tracks based on the total luminosity and $K$ band flux ratio of the binary. Upcoming medium-resolution ($R\approx2000-4000$) JWST/NIRSpec spectroscopy from $1-5~\mu$m (GO3762) should further improve constraints on the bulk properties for Gliese 229 Ba and Bb.

\begin{deluxetable}{lc}[t!]
\tablecaption{Stellar abundances and results of Elf Owl fits \label{table:results}}
\tabletypesize{\small}
\tablehead{
\colhead{Parameter} & \colhead{Value}
}
\startdata
\multicolumn{2}{c}{\textbf{Gliese 229 A}} \\
\hline
C/O\tablenotemark{a} & $0.68\pm0.12$  \\
$\rm{[M/H]}$\tablenotemark{a} & $-0.02\pm0.06$ \\
\hline
\multicolumn{2}{c}{\textbf{Binary brown dwarf model (same abundances)}} \\
\hline
C/O & $0.65\pm0.05$ \\
$\rm{[M/H]}$ & $0.00^{+0.04}_{-0.03}$ \\
$\teffba$ (K) & $900^{+78}_{-29}$ \\
$\teffbb$ (K) & $775^{+20}_{-33}$ \\
$\loggba$ & $5.15^{+0.15}_{-0.04}$ \\
$\loggbb$ & $5.07^{+0.04}_{-0.11}$ \\
$\logkzzbau$  & $3.7^{+1.3}_{-1.4}$ \\
$\logkzzbbu$ & $4.0^{+2.2}_{-1.4}$ \\
$R_1$ ($\Rj$) & $0.81^{+0.05}_{-0.12}$  \\
$R_2$ ($\Rj$) & $0.85^{+0.12}_{-0.05}$ \\
$\lbollsun$\tablenotemark{b} & $-5.19\pm0.01$ \\
$\lbollsunba$ & $-5.40\pm0.01$ \\
$\lbollsunbb$ & $-5.60\pm0.03$ \\
\hline
\multicolumn{2}{c}{\textbf{Single brown dwarf model}} \\
\hline
C/O & $0.65\pm0.05$ \\
$\rm{[M/H]}$ & $-0.02\pm0.04$  \\
$\Teff$ (K) & $840\pm9$ \\
$\logg$ & $4.92\pm0.01$  \\
$\logkzz$ & $3.8\pm0.3$  \\
$R$ ($\Rj$) & $1.165\pm0.015$  \\
$\lbollsun$ & $-5.19\pm0.01$ \\
\enddata
\tablecomments{ We list the median and 95\% credible interval with equal probability above and below the median for parameters derived in this work. These uncertainties only account for statistical error. $_1$ refers to the primary brown dwarf, Ba, and $_2$ refers to the secondary brown dwarf, Bb.}
\tablenotetext{a}{The stellar metallicity is computed from references listed in \S~\ref{sec:abunds}. The stellar C/O is from \citet{Nakajima2015}, and its uncertainty is $1\sigma$.}
\tablenotetext{b}{The luminosities are computed from the measured $\Teff$ and radii from our spectral fits and the Stefan-Boltzmann Law.}
\end{deluxetable}

\subsection{Bolometric luminosity and flux ratio}
We estimate the bolometric luminosity of Gliese 229 Ba and Bb from our model fits using the measured $\Teff$ and radii of each component. We find $\lbollsun=-5.40\pm0.01$ for Ba, and $\lbollsun=-5.60\pm0.03$ for Bb. By summing up the component $\lbol$, we estimate the total luminosity of the combined source to be $\lbollsun=-5.19\pm0.01$, consistent with the value of $-5.21\pm0.05$ estimated using spectro-photometry $<5$ $\mu$m by \citet{Filippazzo2015}. Therefore, we confirm that the under-luminosity problem of Gliese 229 B as a single source is resolved by the binary model.

\begin{figure*}
    \centering
    \includegraphics[width=0.7\linewidth]{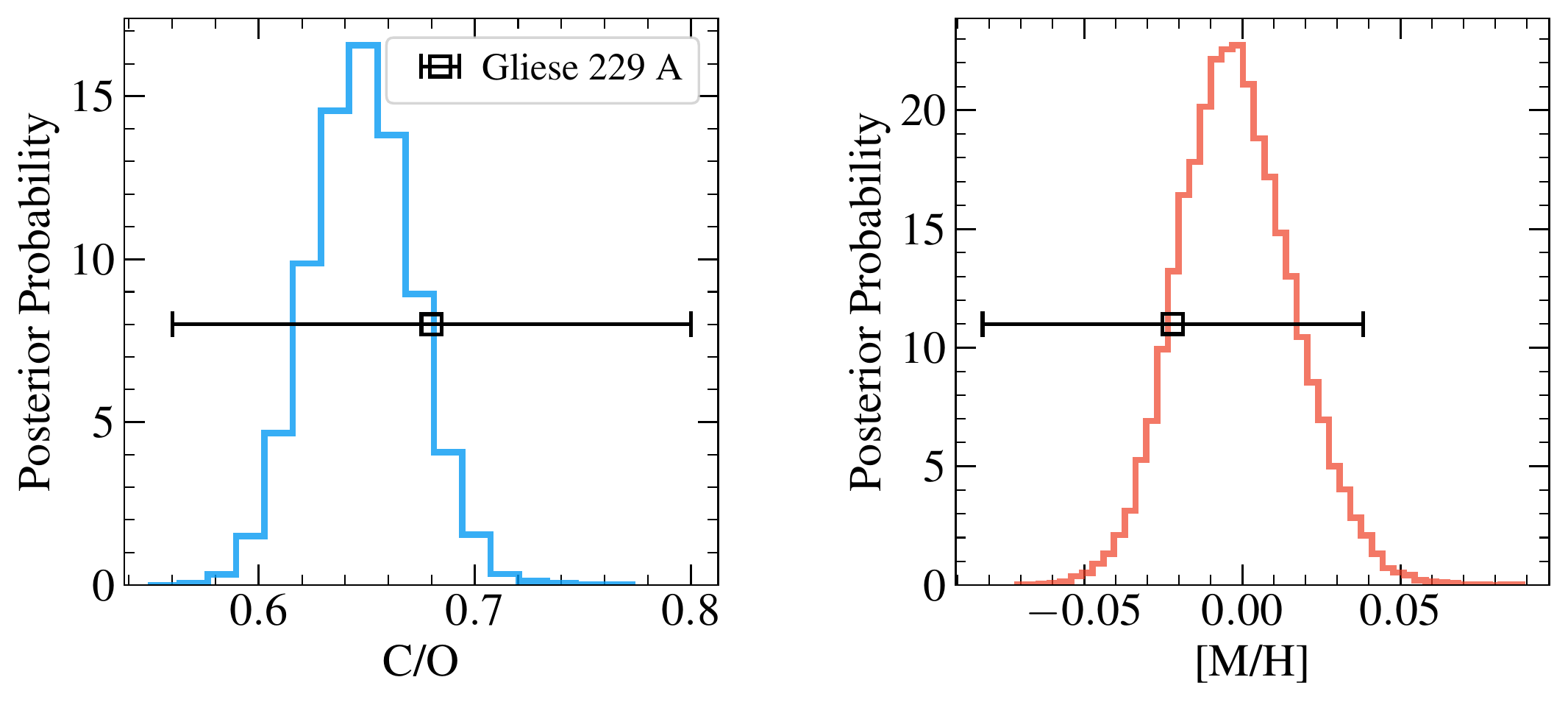}
    \caption{Histograms show the posteriors for C/O and $\rm{[M/H]}$ from our Sonora Elf Owl fits to Gliese 229 Bab. The error bars show the measured abundances for Gliese 229A from the literature, whose sources in given in \S~\ref{sec:abunds}. The abundances of Bab and A are consistent at the $<1\sigma$ level, demonstrating the power of JWST/MIRI LRS in making robust abundance measurements.}
    \label{fig:abund_compare}
\end{figure*}

\subsection{C/O, metallicity, and vertical diffusion parameter}\label{sec:abunds}
We provide updated abundance measurements for Gliese 229 Bab. From our fiducial model which assumes the same abundances for the two brown dwarfs (Table~\ref{table:results}), we find $\rm C/O=0.65\pm0.05$ and $\rm [M/H]=0.00^{+0.04}_{-0.03}$ ($2\sigma$ statistical errors). Alternatively, in our model with distinct C/O and [M/H] values for each brown dwarf, we find $\rm C/O_{Ba} = 0.54^{+0.23}_{-0.14}$ and $\rm C/O_{Bb}=0.82^{+0.29}_{-0.31}$, $\rm [M/H]_{Ba}=0.07^{+0.35}_{-0.18}$ and $\rm [M/H]_{Bb}=-0.07^{+0.23}_{-0.26}$ ($2\sigma$) for the two brown dwarfs (Table~\ref{table:results_app}). These values are consistent at the $1.5\sigma$ and $1\sigma$ levels between Ba and Bb, justifying our assumption of fitting a single set of abundances for the two brown dwarfs. Furthermore, we note that the fiducial model is weakly favored over the model with distinct abundances for Ba and Bb, with log Bayes factor of $1.1$ (or $2\sigma$ preference). 

We note that our reported C/O represents the global or bulk C/O in the atmosphere, as Elf Owl models are parameterized by the global C/O. Therefore, no correction to account for oxygen that is lost to refractory cloud condensation \citep[e.g.][]{Xuan2022, Calamari2024} is needed. The M1V star Gliese 229 A has a C/O measurement from \citet{Nakajima2015}, who used Gemini/IGRINS spectroscopy in $H$ and $K$ bands to derive $\rm C/O=0.68\pm0.12$ for the star, which is slightly higher, but within $1\sigma$ of the solar C/O of $0.59\pm0.08$ from \citet{Asplund2021}. There are numerous metallicity measurements for Gliese 229 A that broadly agree on a near-solar metallicity. Synthesizing the more recent studies \citep{Marfil2021, Rice2020, Gaidos2014,Hojjatpanah2019,Hojjatpanah2020,Maldonado2020,Schweitzer2019,Neves2013,Kuznetsov2019}, we adopt $\rm{[M/H]}=-0.02\pm0.06$ for the star.\footnote{Specifically, we take the weighted average of measurements from these studies and use the standard deviation of the different values as the uncertainty.} Therefore, our values for C/O and $\rm{[M/H]}$ for Gliese 229 Bab are fully consistent with those of its host star (Fig.~\ref{fig:abund_compare}). The chemical homogeneity between Gliese 229 A and Gliese 229 Bab is consistent with expectations from formation via gravitational disk instability or molecular cloud fragmentation. We discuss our new abundance measurements in light of previous studies \S~\ref{sec:discuss}.

\begin{figure}
    \centering
    \includegraphics[width=\linewidth]{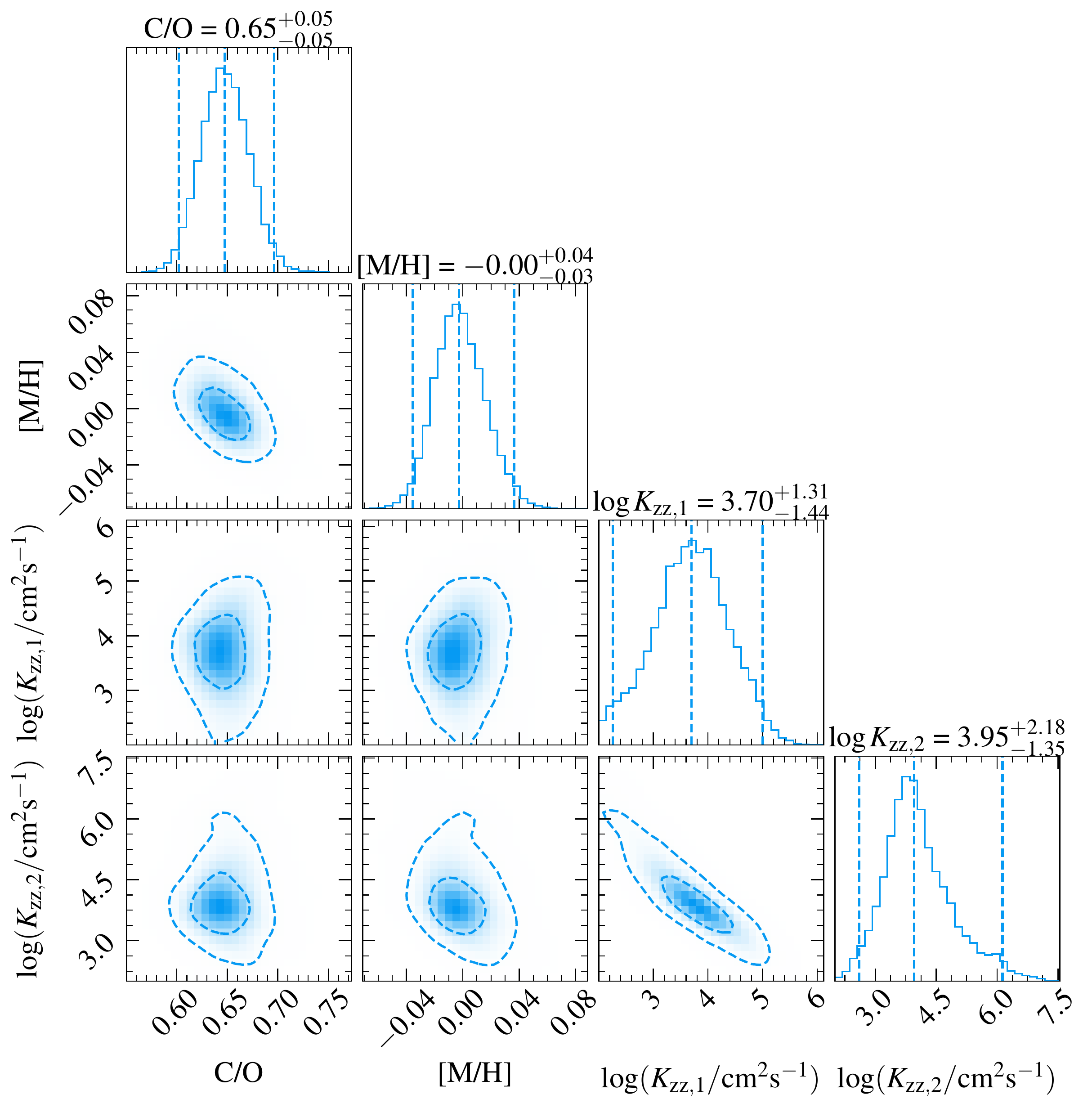}
    \caption{Joint posterior distributions for C/O, $\rm{[M/H]}$, and $\logkzz$ for each brown dwarf from the binary model that assumes the same abundances for Ba and Bb. The dashed lines on the 1D histograms show $95\%$ confidence intervals, while dashed lines on the 2D histograms represent $1\sigma$ and $2\sigma$ contours.}
    \label{fig:corner_ab}
\end{figure}

The strong absorption features of NH$_3$, H$_2$O, and CH$_4$ in our data provide constraints on the vertical diffusion parameter $K_{\rm zz}$. By changing the $K_{\rm zz}$ in otherwise identical models, we observe that higher $K_{\rm zz}$ values quench CO and NH$_3$ at deeper pressures where their abundances are higher, resulting in stronger absorption features for these molecules with higher $K_{\rm zz}$. For the atmospheric conditions of Gliese 229 Ba and Bb, the CO volume-mixing ratio in particular changes strongly with different $K_{\rm zz}$ (a factor of $\approx 3-4$ increase from $\logkzz=2.0$ to $\logkzz=4.0$), which the bluest part ($4.75-5.0~\mu$m) of the MIRI LRS data are sensitive to. 

From our fits, we find clear evidence of disequilibrium chemistry, and measure nearly identical values of $K_{\rm zz}$ for the two brown dwarfs, $\logkzzbau=3.7^{+1.3}_{-1.4}$ and $\logkzzbbu=4.0^{+2.2}_{-1.4}$. These values are generally in line with those found for isolated brown dwarfs of similar $\Teff$ in \citet{Mukherjee2024}.

\subsection{Single brown dwarf fits}\label{sec:single}
As a test, we additionally fit the MIRI LRS data using a single brown dwarf model, whose results are included in Table~\ref{table:results}. Interestingly, we obtain nearly the same abundances from the single brown dwarf model as the binary model, which suggests the binarity of Gliese 229 Bab is not confounding abundance measurements for our MIRI LRS data. This is likely caused by the low-resolution of MIRI LRS, and the high degree of spectral similarity in the mid-infrared between the two brown dwarfs (see Fig.~\ref{fig:nirmodel}). While the binary brown dwarf fit yields slightly lower reduced $\chi^2$ compared to the single fit (1.21 v.s. 1.22), the single brown dwarf fit is statistically favored due to its lower number of parameters, with a log Bayes factor of 4.5 (or $3.5\sigma$ preference). This means the MIRI LRS data alone would have been insufficient to prove that Gliese 229 B is a binary.

The single brown dwarf fit yields $\Teff=840\pm20~K$, which is in between the measured $\Teff$ of Ba and Bb from the binary model. The radius from the single brown dwarf fit is inflated to $R=1.17\pm0.02~\Rj$, which is unusually large for a single object of $71.4~\Mj$ with a field age and another telltale sign of the binary nature of Gliese 229 Bab (see also \citealt{howe_GJ_2022, calamari_Atmospheric_2022}).

\section{Discussion}\label{sec:discuss}

\begin{figure*}[t!]
    \centering
    \includegraphics[width=0.8\linewidth]{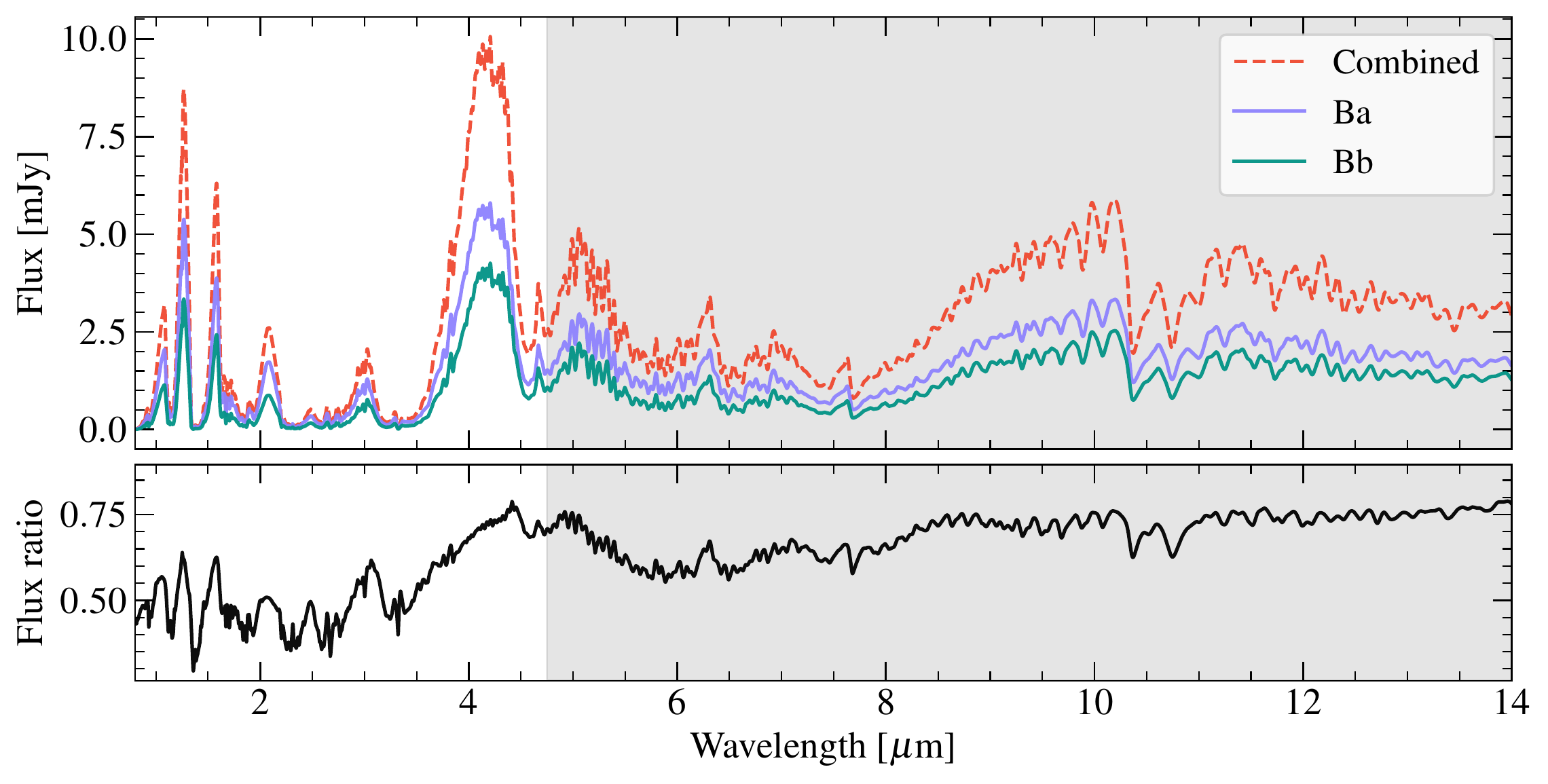}
    \caption{Top panel: the best-fit combined (red dashdot) and component models (light purple and teal) from the binary brown dwarf fit. The models are Gaussian-convolved using constant spectral resolution of 200 for illustration purposes. The shaded grey area indicates the wavelength range covered by the MIRI LRS data. Bottom panel: the computed flux ratio (Bb/Ba) from the best-fit models above. In general, the flux ratio in the near-infrared is lower and shows more structure between the two components.}
    \label{fig:nirmodel}
\end{figure*}

Recently, several studies have noted a trend of super-solar C/O ratios for brown dwarfs (both isolated and bound) from retrieval analyses \citep[e.g.][]{calamari_Atmospheric_2022, Zalesky2022}. For brown dwarfs orbiting stars, the stellar abundances are generally similar to the Sun, which has C/O$=0.59\pm0.08$ \citep{Asplund2021}. Nearby field brown dwarfs in the solar neighborhood are also expected to have broadly solar compositions. This makes the inferred atmospheric C/O of the brown dwarfs, which go up to 1.5, anomalously high in these studies. For Gliese 229 B, two independent retrieval studies have found atmospheric $\rm C/O\approx1.1$ even after accounting for missing oxygen sequestered in condensate clouds \citep{calamari_Atmospheric_2022, howe_GJ_2022}. These values are approximately $3\sigma$ higher than the stellar C/O of $0.68\pm0.12$. Under the assumption that Ba and Bb have the same abundances, our updated measurements of $\rm C/O=0.65\pm0.05$ and $\rm [M/H]=0.00^{+0.04}_{-0.03}$ for Gliese 229 Bab are fully consistent with the abundances of the host star, and unaffected by binarity (\S~\ref{sec:single}). Alternatively, if we fit two distinct sets of C/O and [M/H] for each brown dwarf, we also obtain values consistent with the stellar abundances at $<1\sigma$ level (see Table~\ref{table:results_app}). From our fits, we predict that near-infrared data is more strongly affected by binarity (see Fig.~\ref{fig:nirmodel}), which could potentially bias previous abundance measurements for this particular object. Future work modeling near-infrared spectroscopy of Gliese 229 Bab with two-component models will be needed to examine this hypothesis further. 

Alternative explanations are still required for other brown dwarfs with high C/O measurements, the vast majority of which are not expected to be unresolved binaries based on the low binary fraction in the substellar regime \citep{Fontanive2018}. These explanations could lie in data systematics, modeling issues, or perhaps both \citep{calamari_Atmospheric_2022}. For instance, models could be under-estimating the oxygen abundance. This problem is particularly pertinent for late T dwarfs, which are expected to have silicate clouds below their photospheres. However, quantitatively,  \citet{Calamari2024} found that sequestering oxygen in clouds is insufficient in reconciling the large number of observations that find atmospheric C/O$>0.8$ for brown dwarfs. 

Another potential culprit in the modeling is the use of free chemistry retrievals that assume vertically constant abundances for all species. T dwarfs are known to be in strong chemical disequilibrium, an effect which is self-consistently accounted for in Sonora Elf Owl with the $K_{\rm zz}$ parameter. In addition to free retrievals, it would be informative to run retrievals using chemical equilibrium models along with quenching \citep{Zahnle_methane_2014} to account for vertically-varying abundances and vertical mixing. This approach has been adopted in many studies of directly imaged planets and brown dwarf companions \citep[e.g.][]{molliere_Retrieving_2020, zhang_13COrich_2021, Xuan2022, Inglis2024}, and has generally produced C/O measurements that match the solar or stellar values. Moreover, we demonstrate in this work that the Sonora Elf Owl provides excellent fits to high-quality mid-infrared JWST data for brown dwarfs. This motivates the continued use of forward model fits, in addition to retrievals, in interpreting the atmospheres of brown dwarfs and exoplanets. 

Using different opacities for
molecules such as CH$_4$ could also affect retrieval results. \citet{Hood2023} showed that their retrieved CH$_4$ abundances for a late T dwarf changes when using the newer
and more accurate \citet{hargreaves_Accurate_2020} line list as
compared to the \citet{yurchenko_ExoMol_2014} line list. Specifically, using the newer opacities for CH$_4$ lowers their measured C/O ratio to a solar value. Therefore, using opacities from up-to-date line lists is important in obtaining accurate results. In Elf Owl, CH$_4$ opacities from the \citet{hargreaves_Accurate_2020} line list are used to generate model spectrum.

There could also be unknown systematics on the data side. It is worth noting that elevated C/O measurements in the literature have almost exclusively come from low-resolution ($R\sim100$) near-infrared spectroscopy from $\approx1-2.5~\mu$m. There are known challenges with modeling spectra at these wavelengths, such as theoretical uncertainties in modeling the alkali line wings in the $J$ band \citep{Oreshenko2020, Gonzales2020}, and the increasing impact of cloud opacity at the bluest wavelength. For ground-based observations, it is often necessary to include flux scaling factors between data taken at different epochs and with different instruments. However, because these scaling factors are not known a priori, the results can vary significantly depending on whether scaling factors are included or how priors on these factors are defined \citep[e.g.][]{Xuan2022, Zhang2023}. 

\section{Summary}
In this work, we present an atmospheric analysis of the nearby brown dwarf binary Gliese 229 BaBb using MIRI/LRS spectroscopy ($4.75-14~\mu$m). Using Sonora Elf Owl models, we obtain excellent fits to the data (reduced $\chi^2\approx1.2$). We infer the bulk properties of each brown dwarf. Assuming the two brown dwarfs share the same abundances, we find $\rm C/O=0.65\pm0.05$ and $\rm [M/H]=0.00^{+0.04}_{-0.03}$ ($2\sigma$ credible intervals), which are fully consistent with the stellar values (Fig.~\ref{fig:abund_compare}). We tested an alternative model with different abundances for each brown dwarf, and found this model also yields chemical similarity between the two brown dwarfs and their star.

We find vertical diffusion coefficients of $\logkzz\approx4.0$ for both brown dwarfs, in line with $\logkzz$ measurements for field brown dwarfs of similar \Teff \citep{Mukherjee2024}. Finally, we provide luminosity measurements for both brown dwarfs, and confirm that the total luminosity of the brown dwarf binary ($\lbollsun=-5.19\pm0.01$) is anomalously high for a single object given the measured dynamical mass. 

We discuss our C/O and metallicity measurements in light of previous studies which report anomalously high C/O values for late T dwarfs \citep[e.g.][]{Zalesky2022}, including two retrieval studies on Gliese 229 B \citep{calamari_Atmospheric_2022, howe_GJ_2022} before the object was resolved into a binary by VLTI/GRAVITY and VLT/CRIRES+ \citep{Xuan2024gliese}. Interestingly, the binarity of Gliese 229 Bab does not affect our abundance measurements, as an alternative single brown dwarf model yields the same results. This is potentially due to the very similar $\Teff$ ($\approx 900$ K v.s. $\approx780$ K) of two brown dwarfs. Additional work is required to investigate potential issues with modeling archival near-infrared spectra of T dwarfs, which could be affected by data systematics, modeling uncertainties, or both. The exquisite quality and wavelength coverage of JWST spectroscopy is revolutionizing the data quality, allowing us to perform the most stringent tests of atmospheric and evolutionary models. \\

\vspace{-2mm}
\noindent \textbf{Acknowledgments.}
J.X. thanks Samuel Beiler for discussions about MIRI LRS, Ben Burningham for discussions on opacities and line lists as a source of bias in retrievals, Dino Hsu for discussions about the literature of brown dwarf binaries, and Rebecca Oppenheimer for reading an early draft of this paper. J.X. is supported by the NASA Future Investigators in NASA Earth and Space Science and Technology (FINESST) award \#80NSSC23K1434 and the NASA JWST grant JWST-GO-03762.003-A. Based on observations with the NASA/ESA/CSA \emph{JWST}, obtained at the Space Telescope Science Institute, which is operated by AURA, Inc., under NASA contract NAS 5-03127.
This work benefited from the 2023 Exoplanet Summer Program in the Other Worlds Laboratory (OWL) at the University of California, Santa Cruz, a program funded by the Heising-Simons Foundation and NASA.
The data presented in this article were obtained from the Mikulski Archive for Space Telescopes (MAST) at the Space Telescope Science Institute. 

\vspace{5mm}
\facilities{JWST/MIRI}



\clearpage
\appendix

\FloatBarrier
\section{Custom subtraction of the stellar PSF from the MIRI LRS data}\label{app:data_reduction}
\restartappendixnumbering
Here, we describe the subtraction of the stellar PSF from the data. We began with reductions through the pipeline's Detector1 and Image2 stages. Unlike in default pipeline processing we did not subtract the two nods at this point, deferring that step to later. In other words, this reduction output separate \texttt{cal.fits} files for each of the two dither positions, which is needed for the forward modeling process below. We then performed an additional step to identify and mask outlier pixels: the image was first high pass filtered, and then outliers were identified as pixels that were $>7\sigma$ statistical outliers compared to the estimated uncertainty from the pipeline ERR estimate. The identified pixels were then flagged for exclusion in the data quality array of the original (not high pass filtered) image.

\begin{figure*}[t!]
    \centering
\includegraphics[width=0.95\linewidth]{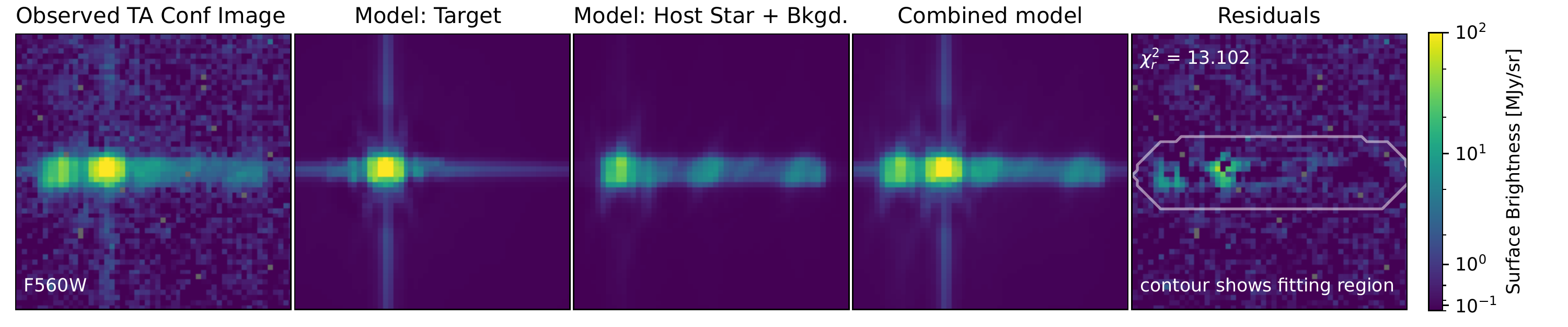}
    \caption{Target acquisition confirmation exposure, and forward modeling. Left panel: The observed TA confirmation image, showing the LRS slit with the companion positioned for the first of two nod positions. Second panel: The best-fit PSF model for the science target. Third panel: The best-fit model for the wings of the host star PSF plus the diffuse flat sky background seen through the LRS slit. The host star Gliese 229 itself is  out of view, $\sim 4.5$ arcsec up and to the left from B. For scale, the LRS slit itself is 4.7 arcsec in length. Fourth panel: The combined forward model summing the previous two panels. Right panel: Residuals of the data minus the model. The goodness of fit metric $\chi^2$ is noted, evaluated over the optimization region indicated by the white contour. This is a reasonably good, though imperfect, fit to this complex scene. The geometric parameters refined in this fit were then used for modeling the LRS-dispersed version of this scene.}
    \label{fig:taconfirm_data_and_model}
\end{figure*}

To remove stellar contamination, we use a custom code\footnote{\url{https://github.com/mperrin/miri_lrs_fm}} to forward model the off-axis host star PSF using WebbPSF \citep{Perrin2012SPIE.8442E..3DP,Perrin2014SPIE.9143E..3XP} and subtract it from the 2D images. The specific version used in this paper can be found in \citep{xuan_miri_zenodo}.\footnote{\url{https://doi.org/10.5281/zenodo.14032760}} A starting estimate for the position of the (mostly unseen) host star relative to the slit can be obtained using the World Coordinate System (WCS) metadata. To refine that estimate for the precise geometry as observed, including compensation for residual errors in FITS header coordinates\footnote{Currently, when target acquisition exposures are used onboard to precisely refine the observatory's pointing, the information about the applied pointing correction is not subsequently used by the ground system and pipeline to improve the astrometric calibration of the science data. In other words, TA exposures improve the \textit{accuracy of the actual achieved pointing} seen in image data, but do not currently improve the \textit{accuracy of the FITS header WCS metadata.}}, we analyze 
the target acquisition verification image observed in the F560W filter. We fit that using a forward model consisting of the planet as a point source seen within the slit, the wings of the offset host star outside of the slit, and the diffuse thermal sky background. We use nonlinear least squares to optimize the precise position offsets and flux scale factors for the companion and host star (see Fig.~\ref{fig:taconfirm_data_and_model}). The best-fit model parameters yield small corrections ($\lesssim 0.25$ MIRI pixels) compared to the WCS.

This modeling incidentally yields a flux estimate for Gliese 229 B in F560W, including correction for slit losses. The measured value is $1.95 \pm 0.2$ mJy. This measurement is not the main purpose of the forward model code and has not been rigorously validated, so we conservatively report 10\% uncertainties for this. Nonetheless the consistency between this F560W photometry and the LRS spectrum at that wavelength range is reassuring. 

\begin{figure*}[t!]
    \centering
\includegraphics[width=0.95\linewidth]{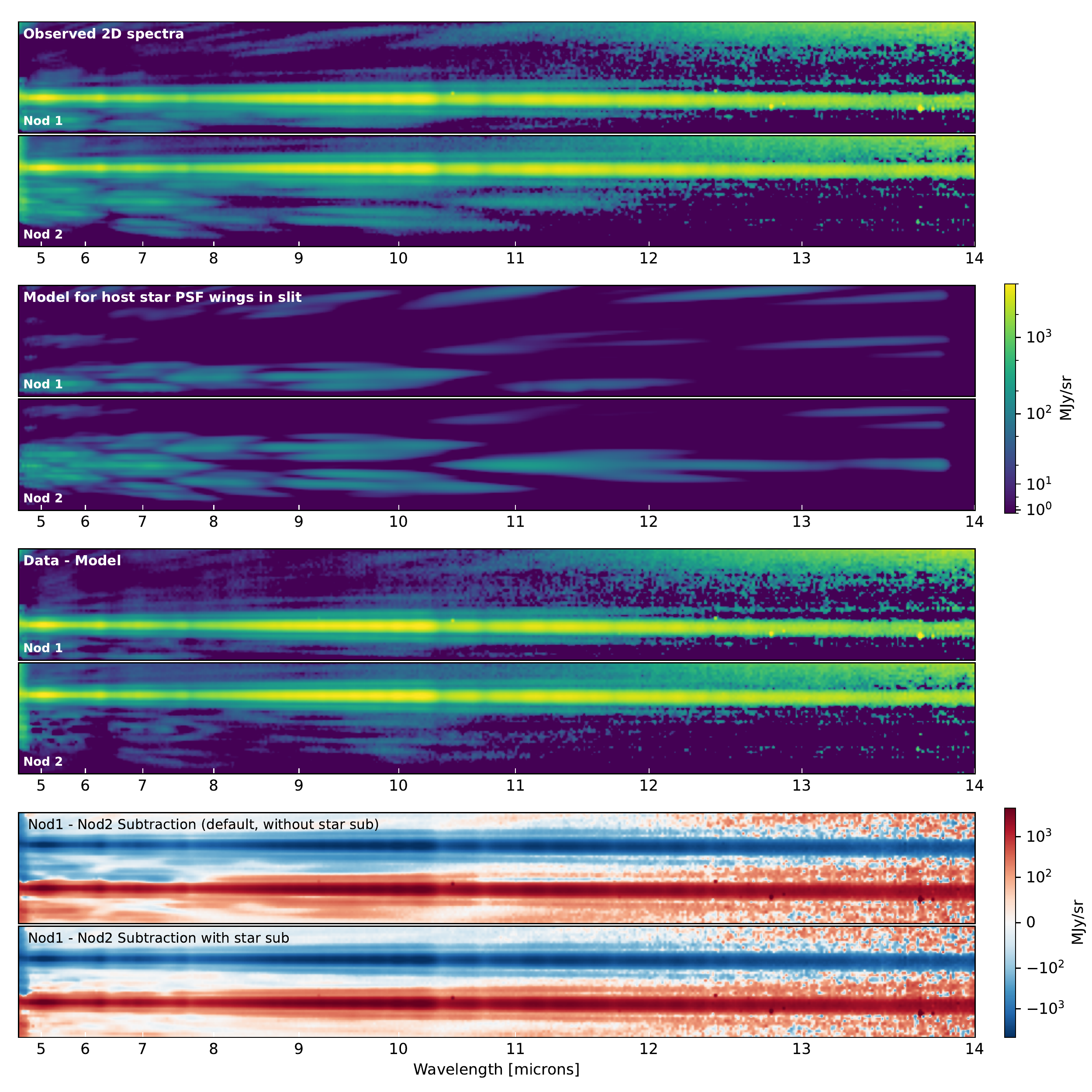}
    \caption{
    \label{fig:lrs_forward_model} Forward modeling of the LRS data to model and subtract the host star light within the slit. \textit{Top 2 panels}: The observed 2D dispersed spectral images for the 2 nod positions, oriented so that the LRS slit direction is vertical and the wavelength dispersion direction is horizontal. These contain both the spectral trace of Gliese 229 B, and also the dispersed PSF wings of the host star. \textit{Second two panels}: Forward model of the dispersed host star PSF wings for the 2 nod positions, generated using WebbPSF as discussed in the text. \textit{Third two panels}: The observed data minus the PSF models. The contamination from the host star PSF wings is significantly reduced, though there are still smaller residuals due to model imperfections. \textit{Bottom panels}: Comparison of nod subtractions without and with the PSF subtraction. In the default subtraction, residual stellar speckes are visible between the two spectral traces, and indeed contaminating on top of those traces. With the star model subtraction, the clean speckle-free regions between the spectral traces show the reduction in contamination.}
\end{figure*}

We then model the host star's PSF wings seen in the dispersed LRS spectral data using that star's inferred coordinates relative to the slit.  Using WebbPSF we generate a series of monochromatic PSFs spanning the LRS spectral range for that offset star. We spatially shifted each monochromatic PSF according to the calibrated spectral dispersion profile of the LRS prism. In other words, we shift each wavelength vertically up or down based on the known wavelength solution of the LRS. 
We sum those monochromatic PSFs to generate a synthetic 2D spectrum, scaling the flux of each wavelength following a model for the host star’s spectral energy distribution. That process yields a 2D forward model of the host star PSF wings as seen through the LRS slit. This process was repeated for the two nod positions in the LRS observation. The resulting models for the dispersed offset stellar PSF in each nod were subtracted from the science data by fitting an overall flux scale factor and a background offset that is allowed to vary linearly with wavelength.

The forward modeling and subtraction of the stellar PSF wings is illustrated in Figure \ref{fig:lrs_forward_model}, which shows that the model provides a clear reduction in the amount of speckle flux in the 2D images. After extracting the 1D spectra (\S~\ref{sec:data_red}, we find that the two nods agree at the $\approx 2-3\%$ level.

\FloatBarrier
\section{Joint posterior distributions from binary model fit}\label{app:corner}
\restartappendixnumbering

\begin{figure}
    \centering
    \includegraphics[width=0.9\linewidth]{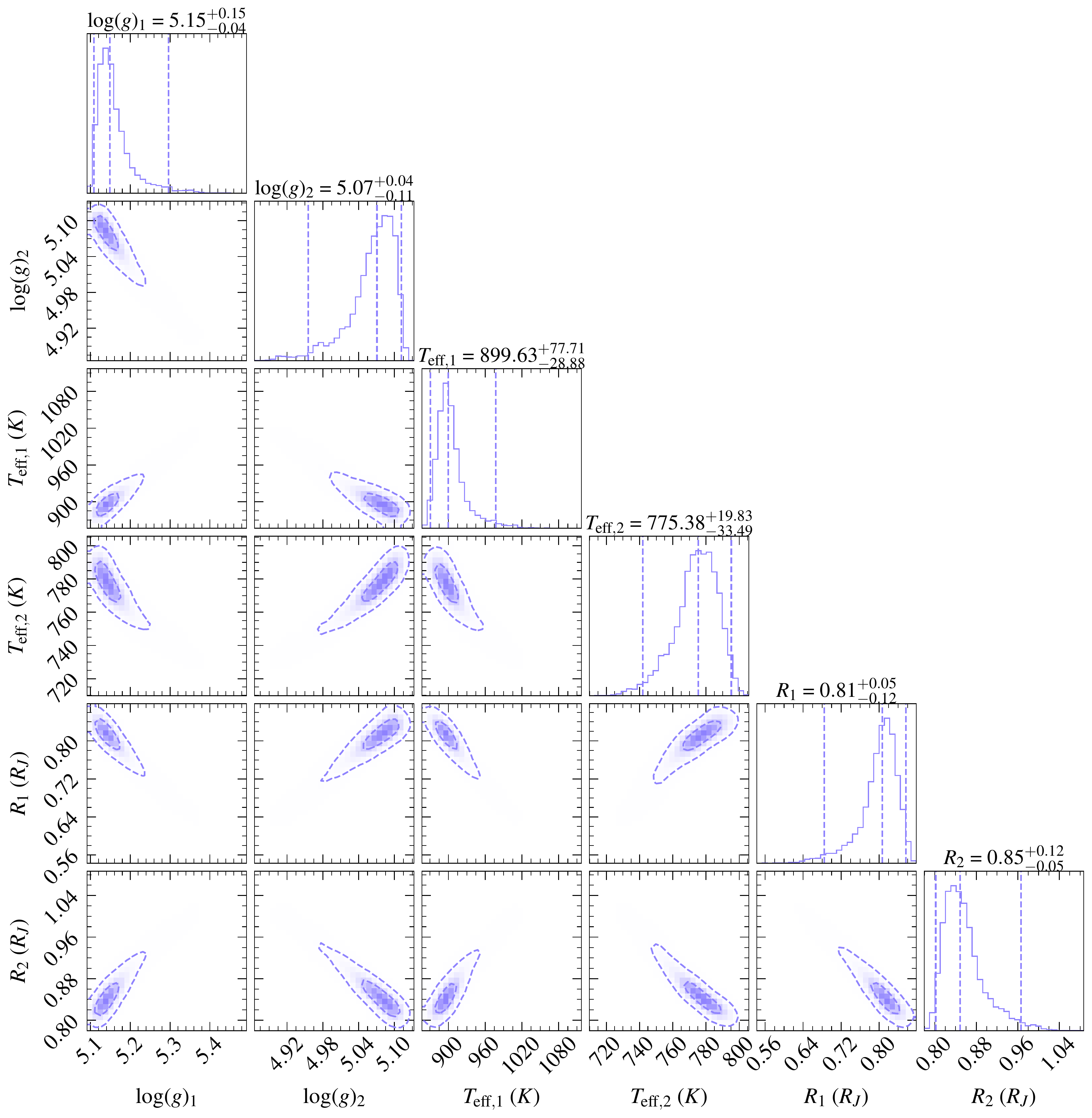}
    \caption{Joint posterior distributions for \logg, \Teff, and radius of each brown dwarf from the binary model. The subscript 1 indicates the primary brown dwarf, Gliese 229 Ba, and 2 indicates Gliese 229 Bb. The dashed lines on the 1D histograms show $95\%$ confidence intervals, while dashed lines on the 2D histograms represent $1\sigma$ and $2\sigma$ contours.}
    \label{fig:corner_bulk}
\end{figure}

Here, we show the joint posterior distributions of bulk parameters from the binary brown dwarf fit in Fig.~\ref{fig:corner_bulk}.

\FloatBarrier
\section{Retrieved resolving power and wavelength correction}\label{app:reslambda}
\begin{figure}
    \centering
    \includegraphics[width=0.45\linewidth]{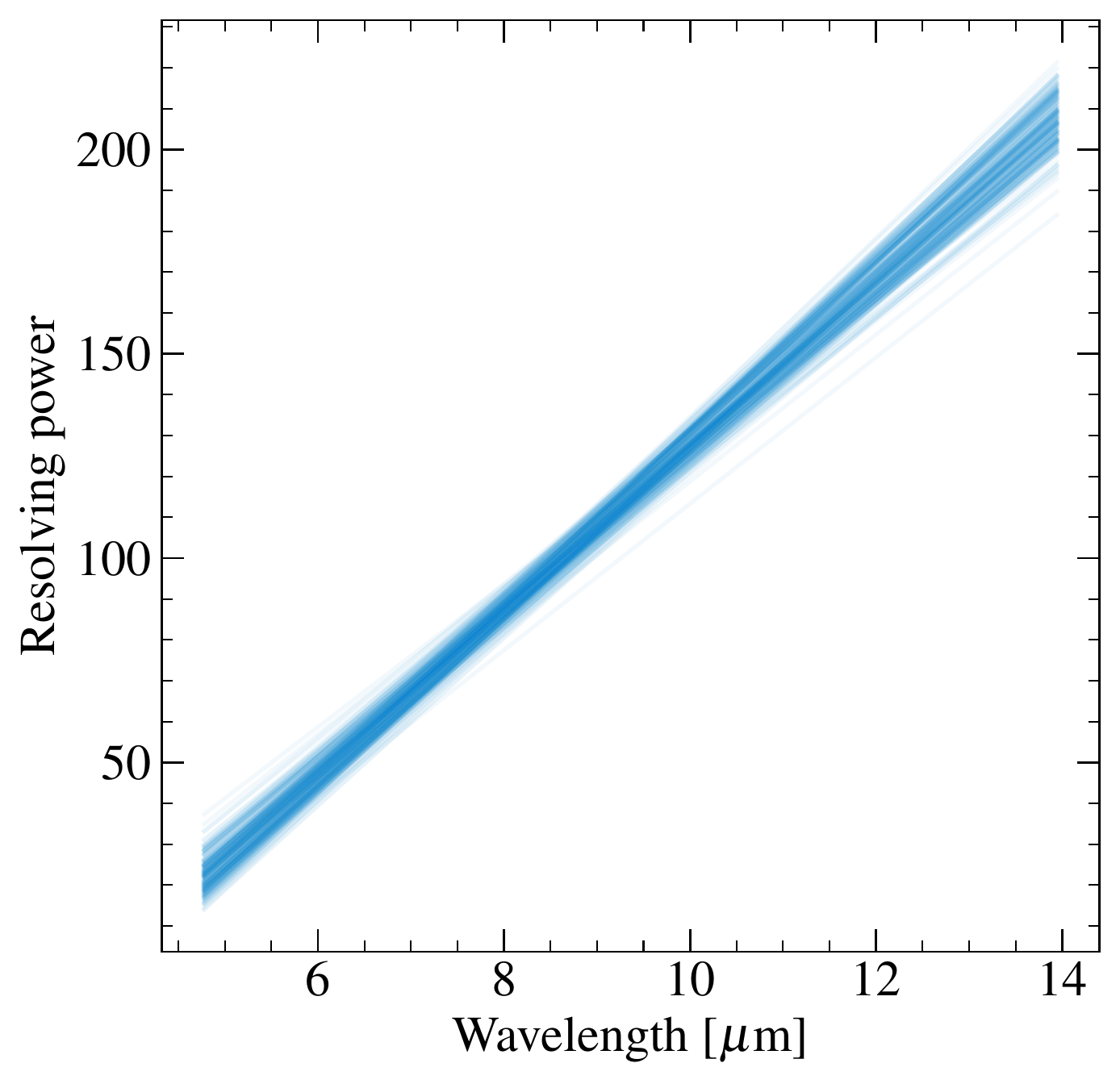}
    \caption{200 random draws of $R\lambda=r_0 + r\lambda$ from an initial single brown dwarf model. }
    \label{fig:rdraws}
\end{figure}

\begin{figure}
    \centering
    \includegraphics[width=0.85\linewidth]{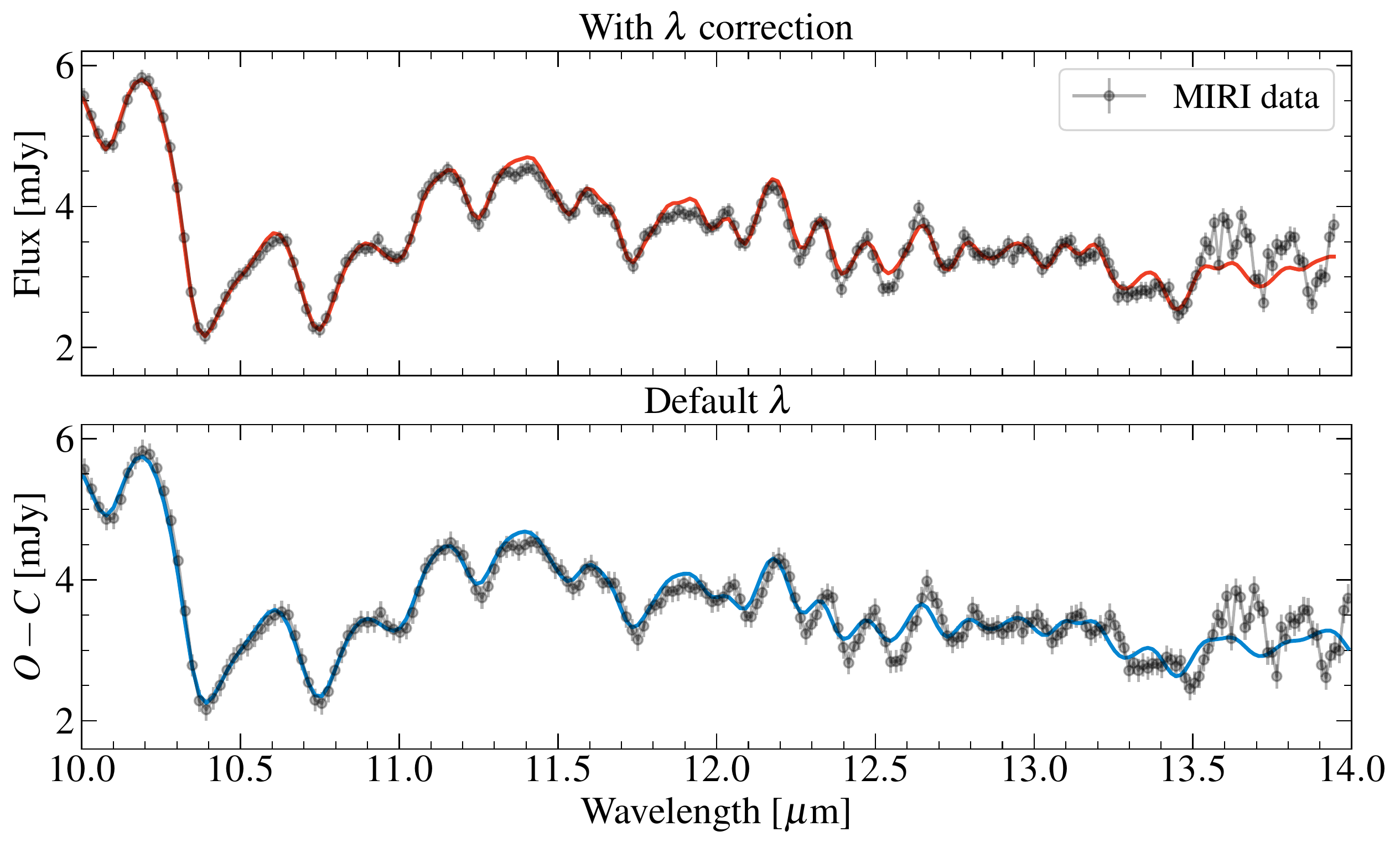}
    \caption{Top: the best-fit single brown dwarf model (red) and MIRI data (black) plotted with the corrected wavelength solution, $\lambda^\prime$. The error bars are inflated by the best-fit error inflation term $b$ from the model. Bottom: best-fit single model (blue) and MIRI data (black) plotted with the default wavelength values from the pipeline. The effect of the wavelength correction is most significant at wavelengths longward of $10\mu$m, which we show here.}
    \label{fig:wcorr}
\end{figure}

\restartappendixnumbering
In Fig.~\ref{fig:rdraws}, we show random draws of the inferred $R_\lambda$ from our single brown dwarf fit. In Fig.~\ref{fig:wcorr}, we compare models and data with and without the wavelength correction of $\lambda^\prime = w_0 + w_1\lambda + w_2\lambda^2$. The best fit parameters are $r=20.0$ and $r_0=-73.1$ for the variable resolving power, and $w_0=-0.0864$, $w_1=1.0223$, and $w_2=-0.0014$ for the wavelength correction. We applied these values in the fits that we report in the paper.

\FloatBarrier
\section{binary brown dwarf fit with two sets of abundances for Ba and Bb}
Here, we provide results of the alternative model, where we fit individual C/O and $\rm [M/H]$ for the two brown dwarfs, along with different $\Teff$, $\logg$, radius, and $\logkzz$. 

\begin{deluxetable}{lc}[H]
\tablecaption{Results of Elf Owl fit for binary model with different abundances \label{table:results_app}}
\tabletypesize{\small}
\tablehead{
\colhead{Parameter} & \colhead{Value}
}
\startdata
\hline
\multicolumn{2}{c}{\textbf{Binary brown dwarf model (different abundances)}} \\
\hline
$\rm C/O_{Ba}$ & $0.54^{+0.23}_{-0.14}$ \\
$\rm C/O_{Bb}$ & $0.82^{+0.29}_{-0.31}$ \\
$\rm{[M/H]_{Ba}}$ & $0.07^{+0.35}_{-0.18}$ \\
$\rm{[M/H]_{Bb}}$ & $-0.07^{+0.23}_{-0.26}$ \\
$\teffba$ (K) & $911^{+65}_{-54}$ \\
$\teffbb$ (K) & $765^{+49}_{-32}$ \\
$\loggba$ & $5.19^{+0.10}_{-0.07}$ \\
$\loggbb$ & $5.03^{+0.06}_{-0.07}$ \\
$\logkzzbau$  & $3.9\pm1.5$ \\
$\logkzzbbu$ & $4.1^{+1.9}_{-1.4}$ \\
$R_1$ ($\Rj$) & $0.77^{+0.06}_{-0.08}$  \\
$R_2$ ($\Rj$) & $0.89^{+0.08}_{-0.06}$ \\
$\lbollsun$\tablenotemark{b} & $-5.18\pm0.02$ \\
$\lbollsunba$ & $-5.40\pm0.09$ \\
$\lbollsunbb$ & $-5.58\pm0.10$ \\
\hline
\enddata
\tablecomments{As for Table~\ref{table:results}, we list $2\sigma$ credible intervals with equal probability above and below for median for these parameters.}
\end{deluxetable}

\clearpage
\bibliography{main.bib}{}
\bibliographystyle{aasjournal}



\end{document}